\documentclass[aps,prb,twocolumn,showpacs,preprintnumbers]{revtex4}
\usepackage{graphics}
\usepackage{graphicx}
\usepackage{dcolumn} 
\usepackage{bm}
\usepackage{color}
\usepackage{epsfig}
\usepackage{amsmath}
\begin{document}
\title{Organometallic-like localization of 4d-derived spins\\ 
in an inorganic conducting niobium suboxide}
\author{K.-W. Lee$^{1,2}$}
\email{mckwan@korea.ac.kr}
\author{W. E. Pickett$^{3}$}
\email{pickett@physics.ucdavis.edu}
\affiliation{
 $^1$Department of Applied Physics, Graduate School, Korea University, Sejong 339-700, Korea\\
 $^2$Department of Display and Semiconductor Physics, 
  Korea University, Sejong 339-700, Korea\\
 $^3$Department of Physics, University of California, Davis,
  CA 95616, USA
}
\date{\today}
\pacs{71.27.+a,71.20.Be,72.15.Qm,75.20.Hr}

\begin{abstract}
Based on the refined crystal structure
comprised of columns of 3$\times$4 planar blocks of NbO$_6$ octahedra
and first principles electronic structure methods,
we find that orthorhombic ($o$)-Nb$_{12}$O$_{29}$ introduces a new class of transition 
metal oxide.  The electronic system consists of a large
Nb dimer-based localized orbital comparable in size to those in organometallic
compounds, yet is tightly bound and weakly interacting with itinerant
electronic bands. These local moments -- a rare occurrence for Nb
-- form one-dimensional spin chains that criss-cross
perpendicularly oriented conducting ``nanowires.''
The local moment bandwidth is comparable to what is
seen in rare earth compounds with extremely localized orbitals. The microscopic origin
is traced to the local structure
of the NbO$_6$ octahedra and associated orbital+spin ordering. The resulting
  1D$_s$$\times$1D$_c$ 
anisotropic two dimensional Heisenberg-Kondo lattice model 
  ($s$=spin, $c$=charge)
provides a strongly anisotropic spin-fermion lattice system
for further study.
\end{abstract}
\maketitle
\vskip 0.7in

\section{Introduction}
Transition metal oxides form some of the most intellectually rich
electronic phases, {\it viz.} the high temperature superconducting cuprates,
the colossal magnetoresistance manganites, and ordering of charge, spin,
orbital and structural changes, especially in $3d$ oxides.
Niobates are $4d$ oxides that assume a fascinating sequence of crystal
structures with assorted properties, extending from the cubic
one-to-one, formally Nb$^{2+}$, metal NbO to the Nb$^{5+}$-based wide-gap insulator
Nb$_2$O$_5$, and encompassing a sequence of mixed valent compounds
such as Nb$_{12}$O$_{29}$,
Nb$_{22}$O$_{54}$, Nb$_{25}$O$_{62}$, Nb$_{47}$O$_{116}$, and so on.

The 12-29 member, Nb$_{12}$O$_{29}$, displays local moment behavior
coexisting with a relatively
high density of conducting
carriers.\cite{cava1991}
Just how local moments -- very rare for Nb -- emerge and coexist with itinerant
electrons in this system has remained an enigma for decades.
More broadly, systems of local moments embedded in conducting media form a rich platform
for unusual phases, with phenomena including Kondo systems, heavy fermion
metals and superconductors, and still unexplained non-Fermi
liquid behavior. 

Nb$_{12}$O$_{29}$ originally attracted attention\cite{iijima1975,meyer1982}
due to the mixed valent
character it implied for Nb: 2 Nb$^{4+}$ + 10 Nb$^{5+}$ balancing the oxygen
charge, apparently leaving  two Nb cations with $4d^1$ configuration. 
It lies just inside the line of conducting compounds in the Nb$_2$O$_{5-2x}$
system, where coexistence of polarons and bipolarons had provided the  
prevailing picture of its conducting behavior.\cite{Ruscher1988,Ruscher1991} 
The tungsten bronzes, which
share the same underlying WO$_3$ structural motif, have been found to 
superconduct up to 4 K when heavily doped.\cite{Sweedler1965,Haldo2014}

Nb$_{12}$O$_{29}$ is found in two polymorphs, orthorhombic ($o$-) and
monoclinic ($m-$), both based on stacking of the same underlying structural
feature (described below) but being stacked differently. The magnetic susceptibility
of each polymorph displays Curie-Weiss susceptibility corresponding to one
spin-half moment per f.u.\cite{cava1991} $m$-Nb$_{12}$O$_{29}$ orders around 12 K,
in the sense that the susceptibility follows a behavior that peaks before dropping, 
in a manner that can be fit by the Bonner-Fisher form for a one dimensional
(1D) Heisenberg antiferromagnet.\cite{lappas2002} 
$o$-Nb$_{12}$O$_{29}$ does not show such a 
peak down to 2 K.\cite{andersen2005}    

This first (and still one of few) observation of Nb local moment 
behavior, moreover to be coexisting
with Nb conductivity,\cite{Ruscher1991,cava1991} suggested a
more specific picture\cite{cava1991} of 
one $d^1$ electron becoming localized and magnetic while the other is itinerant
and Pauli paramagnetic.
While this compound is a very bad metal, it is nevertheless also a standout 
transparent conductor, displaying resistivity $\sim$~3~
m$\Omega$ cm with only 10\% variation\cite{cava1991,ohsawa2011} 
from room temperature to 2 K, 
and it possesses a high carrier
density\cite{ohsawa2011} of one carrier per Nb thus not requiring tuning via 
extrinsic doping.
Fang {\it et al.} recently argued from a density functional study using
the generalized gradient approximation (GGA) functional that the monoclinic
polymorph is an itinerant Stoner magnet.\cite{Fang} Our studies provide a very different
picture of $o$-Nb$_{12}$O$_{29}$, and we contrast the two pictures below.

\begin{figure}[tbp]
{\resizebox{7.5cm}{5.0cm}{\includegraphics{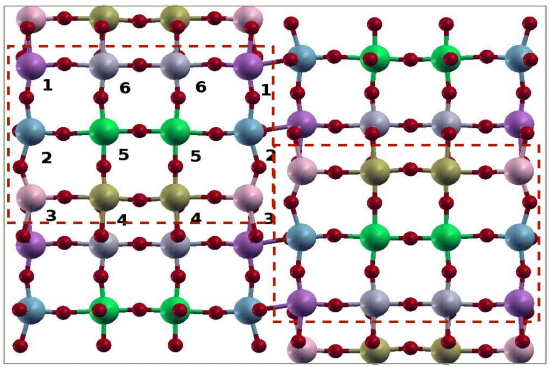}}}
\caption{(Color online) View of the structure in the $\hat c - \hat b$ plane
($\hat c$ horizontal) of $o$-Nb$_{12}$O$_{29}$. 
The designations of the sites Nb1, Nb2, ..., Nb6
in the $3\times 4$ unit are indicated. The dashed lines outline two of the 
3$\times$4 units 
that lie on the same level. Other such units (only partially visible) lie 
at a height $\frac{a}{2}$ along $\hat a$.
Note that the (green) Nb5 site is the only one not on a
`surface' of the 3$\times$4 unit.
}
\label{Structure}
\end{figure}

In this paper we demonstrate that orthorhombic $o$-Nb$_{12}$O$_{29}$ 
presents an unprecedented type of material, based on an unusual large
but strongly localized and magnetic organometallic-like orbital.\cite{O-M2006}
This compound consists at
the most basic level of an array of
linear (1D) Heisenberg magnetic chains along the one direction, crisscrossed by
1D conducting nanowires along a perpendicular direction, coupled to a two dimensional
(2D)  system by Kondo interaction
at each site. This new 1D$\times$1D Heisenberg-Kondo
lattice (HKL) system arises from the
intricate structure of columns of 3$\times$4 planar units of NbO$_6$ octahedra, with
the spin-half {\it dimer Mott insulating system} arising from quantum confinement induced
extreme localization on Nb {\it dimers} with a specific
$4d$ orbital orientation, while other Nb sites provide electronic conduction.
These results should stimulate further study both experimentally, to identify the
very low temperature and high magnetic field phases, and theoretically to probe
a new model system for novel magnetic and possibly exotic superconducting
states, since related 2D models have been shown to harbor pairing correlations.

\section{Structure and Approach}
Nb$_{12}$O$_{29}$ assumes two polymorphs, 
monoclinic ($m$-)\cite{mcqueen2007,waldron2001,lappas2002,waldron2004,andersen2005,cheng2009} 
and orthorhombic ($o$-).\cite{cava1991,andersen2005,mcqueen2007}
Each is based on the same 3$\times$4 block structure of corner-linked
NbO$_6$ octahedra (ReO$_3$-type perovskite) that are stacked along the
$\hat a$ direction with lattice constant $a$ =
3.832~\AA, thus forming rectangular $\infty \times$3$\times$4 nanocolumns
(in $\hat a,\hat b,\hat c$ order of the Nb sites). The 3$\times$4 blocks contain six
crystallographic sites Nb1, ..., Nb6.
The two structures differ only in how these columns are repeated in the
$\hat b-\hat c$ plane, a common motif in polymorphic shear structures.

We focus here on $o$-Nb$_{12}$O$_{29}$\cite{cava1991,andersen2005,mcqueen2007}
with $b$=20.740~\AA, $c$=28.890~\AA,
whose crystal structure was only
settled by use of xray diffraction together with convergent beam electron diffraction
on single crystals by McQueen {\it et al.}\cite{mcqueen2007}
The structure 
at 200 K, shown in Fig.~\ref{Structure}, 
has centered orthorhombic space group $Cmcm$ (\# 63) with two f.u.  per primitive cell.
The primitive cell consists of two such columns, one of which lies
$\frac{a}{2}$ lower/higher than the other, so
these two NbO$_6$ columns are connected edgewise rather than corner-linked.
These shear boundaries, we find,
severely limit carrier hopping in the $\hat b-\hat c$ plane.  Due to the distortion of the Nb-O bond
lengths and angles, the 3$\times$4 unit is ferroelectric
in symmetry. The
centering+reflection symmetry operation however results in no net 
electric polarization in the cell.

Density functional theory (DFT) calculations were carried out
initially using the spin-polarized Perdew-Burke-Ernzerhof GGA functional,\cite{PBE}
implemented in an all-electron full-potential code {\sc fplo}.\cite{fplo}
The on-site Coulomb repulsion $U$ was treated within the
the GGA+U approach.\cite{LDAU,Erik}
We present results for $U=3$ eV, the value calculated\cite{Vaugier} for Nb in perovskite
SrNbO$_3$, and Hund's energy $J^H=1$ eV;
we verified that results are insensitive to the specific values. 
A similar value of $U$ has been applied in the DFT 
plus dynamical mean field method to 
study the dynamical excitation spectrum of Li$_{1-x}$NbO$_2$,\cite{linbo2} which as
one of the few strongly correlated niobates has become of interest because as a
strongly layered transition metal oxide it becomes superconducting, thus displaying
several similarities to the high temperature superconducting cuprates.  
The Brillouin zone was sampled with a regular fine mesh up to
657 irreducible $k$ points.

\begin{figure*}[tbp]
{\resizebox{11.5cm}{8.5cm}{\includegraphics{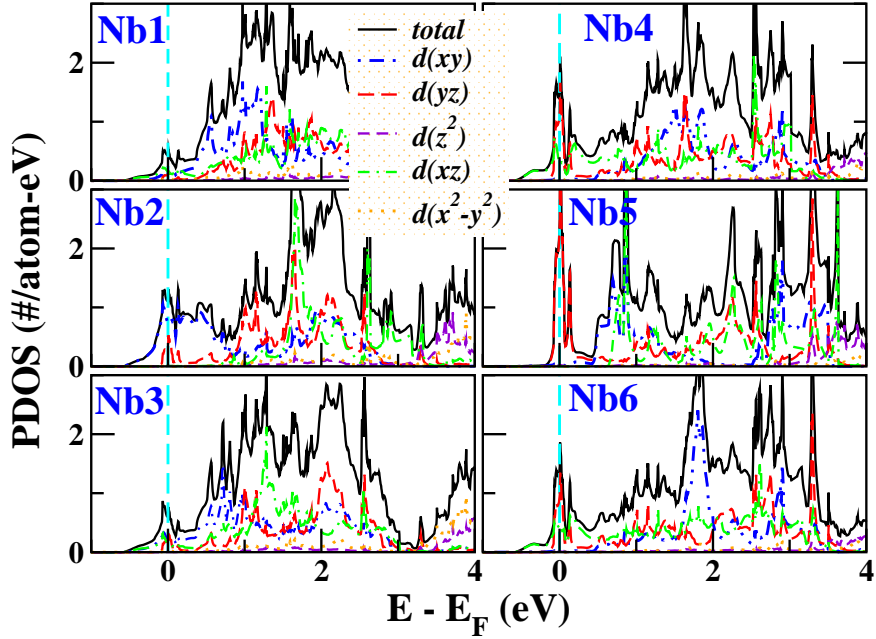}}}
\caption{(Color online) Total and orbital-projected density of states for each
of the Nb sites in nonmagnetic Nb$_{12}$O$_{29}$. 
Note that Nb5 dominates the peak at the Fermi level E$_F$, 
while Nb2 dominates the occupied states at and below E$_F$.
}
\label{PDOS}
\end{figure*}

\begin{figure}[tbp]
\vskip 4mm
{\resizebox{7.5cm}{5.0cm}{\includegraphics{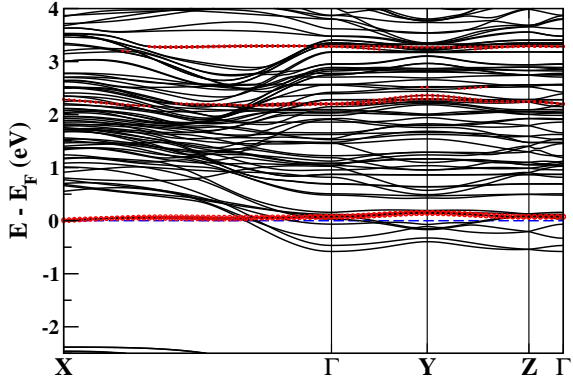}}}
\caption{(Color online) The full band region of non-magnetic Nb$_{12}$O$_{29}$,
above the gap separating them from the O $2p$ bands. 
The red horizontal lines indicate a single, remarkably flat,
$d_{yz}$ bonding and (split apart) antibonding band pair of the Nb5$_2$ dimer.
Symmetry points are X=$(\frac{\pi}{a},0,0),$ Y=$(0,\frac{\pi}{b},0),$
Z=$(0,0,\frac{\pi}{c}).$
}
\label{pmBands}
\end{figure}

\section{The Theoretical Picture}
\subsection{Electronic configuration}
We first address the electronic structure before consideration of magnetism. The system
was earlier found to be too intricate for H\"uckel methods to elucidate.\cite{canadell2000} The atom-projected
density of states (PDOS) in Fig.~\ref{PDOS} reveals a sharp and narrow
Nb5-dominated peak, with secondary contribution from
Nb4 and Nb6, that pins the Fermi level E$_F$ and will be found to be unstable
to local moment formation. Itinerant band states around E$_F$ are dominated by
the Nb2 site, with smaller Nb3 participation but {\it none}
from Nb5.
The band structure in Fig.~\ref{pmBands} reveals a 
Nb5$_2$ {\it dimer bonding band}
lying within the continuum of $\hat a$-axis dispersive conduction bands, 
with remarkable  flat-band dispersion of no more than 75 meV along
each of the three axes.  The antibonding partner lies $\sim$2.8 eV higher and is
bifurcated by additional interactions.
From the small dispersion of the flat Nb5 band, we obtain the nearest
neighbor hopping amplitudes
along the three axes: $t_5^a \approx -10$ meV, $t_5^b = 17$ meV, $t_5^c = 0$.
(The dispersion along $\hat a$ is however not simple nearest neighbor type, in
spite of the small value.)

The anomalously flat band arises primarily from the Nb5 $d_{yz}$ orbital, 
which significantly has its lobes lying in the $\hat b - \hat c$ plane, perpendicular
to the column $\hat a$ direction.
This orientation provides only a tiny $dd\delta$
overlap ($t_5^a$ above) for hopping along the short $\hat a$ axis. The hopping amplitudes given above 
are orders of magnitude smaller than the on-site Coulomb repulsion that will produce Mott
localization of $d_{yz}$ states on the Nb5$_2$ dimer.  The two neighboring Nb5 $d_{yz}$ 
orbitals form
bonding-antibonding combinations by strong $dd\pi$ coupling, splitting the molecular orbitals
by $2 t_{5,\pi}$ = 2.8 eV. The Nb5$_2$ dimer bonding combination $\phi_b$
is partially filled and pins $E_F$. 

\begin{figure}[tbp]
\vskip 4mm
{\resizebox{7.5cm}{5.0cm}{\includegraphics{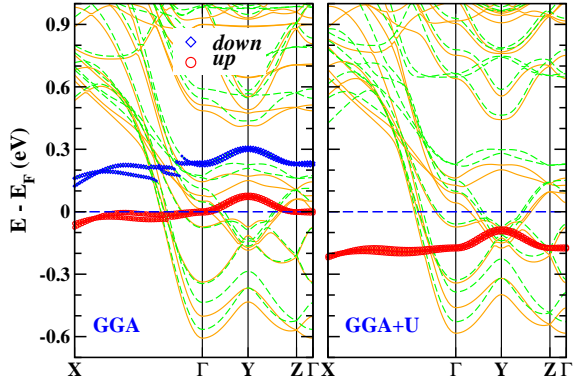}}}
\caption{(Color online) Fatband plots of the bands of Nb$_{12}$O$_{29}$ 
near $E_F$, without and with the strong interaction ($U$).
The magnetic bands within GGA (left panel) highlight
the flat Nb5 $d_{yz}$ band; the dispersive bands within $\pm$0.6 eV of E$_F$
are strongly Nb2 $d_{xy}$-derived.
The right panel shows the bands from a
ferromagnetic GGA+U calculation, with lower Hubbard band in red.
The upper Hubbard band lies above this energy window.
The splitting of majority (resp. minority) conduction bands (yellow, resp. green),
$\sim$40 meV near E$_F$, provides a measure of the Kondo coupling
of the spins to the itinerant electrons.
}
\label{fatBands}
\end{figure}

The Nb2-derived conducting ``wires'', a pair on opposing columnar faces perpendicular to $\hat c$
and two columns per primitive cell, lead to four bands that disperse through E$_F$ along 
($k_a$,0,0).
Coupling between the conducting wires lifts the degeneracy, 
as can be seen from the occupied bands
along $\Gamma$-X in Figs.~\ref{pmBands} and \ref{fatBands}.
Only two bands cross E$_F$ along $\Gamma$-Y. 
There are four flat, 1D sheets perpendicular to $\hat a$, plus some small and probably 
unimportant sheets. 
These flat sheets, with wavevectors
$k_F$$\approx$$\pm (0.15-0.35)\frac{\pi}{a}$, imply large susceptibilities and impending 
intra-columnar 1D instabilities
at several Fermi surface nesting calipers.  The several different nesting wavevectors that connect the
various 1D FS sheets will however tend to frustrate any specific
instability ({\it viz.} charge or spin density waves, or Peierls distortions).

\begin{figure}[tbp]
{\resizebox{5.0cm}{3.5cm}{\includegraphics{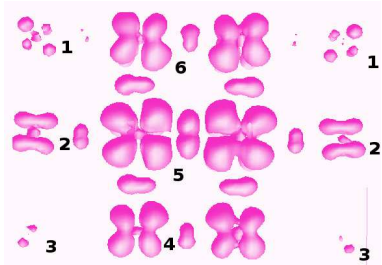}}}
\caption{(Color online) Nb5$_2$-centered spin density isosurface, 
looking down on one layer of the 3$\times$4 unit in the $\hat b - \hat c$ plane, 
of the magnetic electron in Nb$_{12}$O$_{29}$. 
Since the state is weakly mixed in all three direction,
the spin density is also the density of the corresponding Nb5$_2$ 
dimer centered molecular orbital $\phi_b$.
A majority of the density lies on the central Nb5$_2$ dimer $d_{yz}$
orbitals and the intervening O $p_y$ orbital.
}
\label{SpinDens}
\end{figure}

\subsection{Magnetic character}
The conventional GGA treatment of electronic
exchange and correlation processes  
produces a  Slater magnetic moment 
arising from the exchange-split flat band. However,
in flat-band transition metal oxides such as this (though rarely as extreme), 
explicit attention to on-site
repulsion is important, so we focus on the GGA+U treatment\cite{LDAU,Erik}
that is often successful in treating strong interaction effects.
Here $U$ is the Hubbard repulsion
on the Nb ions supplementing the semi-local GGA exchange-correlation
potential.\cite{PBE}

From the correlated treatment,
one peculiar feature is that the magnetic moment arises from a single electron
on the Nb5$_2$ {\it dimer} rather than on a single atom. 
This distinction makes it a quarter-filled band system that results in
a dimer Mott insulating (dMI) subsystem,\cite{dMI} a phase not yet reported in 
otherwise conducting 3D systems. Dimer Mott insulators are rare, although a
{\it trimer} Mott insulator has been reported.\cite{trimerMI} 
The Nb5$_2$ dimer orbital density $|\phi_b|^2$, also the spin density, is displayed
in Fig.~\ref{SpinDens}, and its bonding character (incorporating the intervening
O ion) is evident.

This dMI occupied state however lies
within the continuum of occupied conduction bands.  From the nearly invisible band
mixing with the conduction bands (see Fig.~\ref{fatBands}), it is found that
band mixing (charge coupling) with the conduction
electrons is quite small. Then, from the splitting of majority and minority conduction
bands (see Fig.~\ref{fatBands}) the exchange (Kondo) coupling along $\hat b$, 
$J_K \vec s_j \cdot \vec S_j$, between itinerant $\vec s_j$ and localized $\vec S_j$ spins
at site $j$, has coupling strength $J_K$ = 20 meV that reproduces the spin splitting
of the conduction bands in Fig.~\ref{fatBands} for ferromagnetic alignment. 

Three features may be contributing to the special behavior of the Nb5 site. First, 
quantum confinement: Nb5 is the only site interior to the 3$\times$4 block, it is
not an interface atom. Second, its octahedron is much more regular
than for the other sites, having only a single Nb5-O distance\cite{mcqueen2007} greater than 
2.00~\AA.
Third, the anisotropic displacement parameter ($U_{11}$ describing displacement  from
the ideal position in the $\hat a$ direction) is six times larger than for any other
Nb site, reflecting dynamic (or possibly spatial) fluctuations in position. 

\begin{figure}[tbp]
{\resizebox{6.0cm}{4.2cm}{\includegraphics{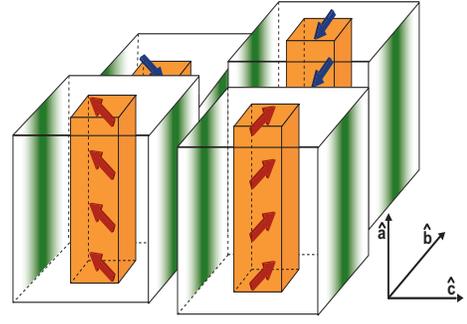}}}
\caption{(Color online) Schematic representation of the active states in
 Nb$_{12}$O$_{29}$. Spin chains (arrows) form columns along $\hat a$ 
but are coupled most strongly along $\hat b$ (hence shown antialigned),
while they are uncoupled along $\hat c$.
Each spin is positioned
between two `wires' conducting along the $\hat a$ axis; the green
shading gives an idea of the conduction electron density spread.
}
\label{Schematic}
\end{figure}

Nb$_{12}$O$_{29}$ thus presents a novel type of system with interacting local moments
in a conducting background -- a Heisenberg-Kondo lattice system. The earlier 1D HKL model
has been generalized to two-dimensions in a maximally anisotropic matter: spin coupling
in one direction and fermion hopping in the perpendicular direction.
The basic picture is illustrated schematically in Fig.~\ref{Schematic}.
Geometrically, the picture is one of 1D Heisenberg spin chains of Nb5$_2$ local moments
extending along $\hat b$,
sandwiched between nanowires (or ``tapes'') conducting along $\hat a$, lying on the 
faces of the columns perpendicular to $\hat b$.
The lack of dispersion along $\hat c$ can be ascribed both to the electronic disruption
caused by the
shear boundaries, and to the centering+mirror symmetry operation that serves to 
produce maximum separation of neighboring Nb sites.

This picture of $o$-Nb$_{12}$O$_{29}$ is differs from that of $m$-Nb$_{12}$O$_{29}$
presented by Fang {\it et al}.\cite{Fang} Based on GGA alone and the VASP code,
they presented a picture of a weak itinerant (Stoner) moment 
arising from a flat
band with magnetic exchange splitting of 0.3 eV.  This result is much like what we obtained
for $o$-Nb$_{12}$O$_{29}$ at the GGA level (see Fig.~\ref{fatBands}). Our view is that
the flat band requires consideration of correlation effects. 
Several specific differences
follow, viz. strong local moment leading to a dMI subsystem
versus a spin-split band Stoner moment in a fully itinerant system, and the
corresponding coupling to the carriers. Such differences can be resolved by
more extensive experimental study at low temperature.

\section{Discussion}
From the hopping amplitudes given above, the largest
magnetic coupling is not along the structural $\hat a$ chain, 
but rather between neighboring Nb5$_2$ 
pairs in the $\hat b$ direction,
corresponding to a superexchange coupling through the oxide ions at the edges
of the 3$\times$4 blocks of $J_H = 4(t_5^b)^2/U \sim 0.5$ meV = 6 K.
We presume the direct exchange will be smaller, based on the molecular orbital
pictured in Fig. 5, because the direct overlap with neighboring orbitals seems very minor.
The moments will be relatively uncorrelated along the conducting $\hat a$ direction
($J^a \sim 2$ K), with the dominant
antiferromagnetic (AFM) correlations developing along $\hat b$ only at low temperature below $J^b$. This temperature
scale is roughly consistent with that from susceptibility $\chi$(T),\cite{andersen2005}
which shows spin-half behavior with
$\theta_{CW}$ = -14 K, but no deviation from Curie-Weiss behavior
down to 2 K. The monoclinic phase,
which we will address elsewhere, ``orders'' via displaying a $\chi$ peak at 12 K that
can be fit\cite{lappas2002} with the Bonner-Fisher form\cite{Bonner} 
for the 1D Heisenberg chain (which shows no true long-range
order). 

The basic Hamiltonian to model such a system, neglecting band indices and spin chain
indices for the simplest model, is
\begin{eqnarray}
H &=& -t \sum_{j,\alpha} [c^{\dag}_{j,\alpha}c_{j+{\hat a},\alpha} + 
                          c^{\dag}_{j,\alpha}c_{j-{\hat a},\alpha}] \nonumber \\
  &+&  J_H\sum_j \vec S_j \cdot \vec S_{j+{\hat b}}
  + J_K \sum_{j} \vec s_j \cdot \vec S_j,
\end{eqnarray}
where $j,\alpha$ run over the 2D lattice and spin directions, 
$\hat a$ and $\hat b$ are the 2D lattice vectors,
$\vec \sigma$ is the Pauli spin matrix vector, and
$\sum_{\alpha\beta} c^{\dag}_{j,\alpha} \vec \sigma_{\alpha\beta} c_{j,\beta}=\vec s_j$
is the conduction electron spin at site $j$ in terms of the carrier
creation operator $c^{\dag}_{j\sigma}$.
The energy scales are $t$=400 meV, $J_H$$\approx$0.5 meV, $J_K$=20 meV.
In this model an itinerant electron is confined to its nanowire, and suffers only Kondo (spin)
scattering due to the interaction. The 1D Heisenberg chains likewise are perturbed
only by spin coupling to itinerant electrons.  

The strictly 1D version of the Heisenberg-Kondo lattice model has been studied 
actively\cite{Zachar1996,Berg2010}
in the context of high temperature superconducting cuprates, where 1D structures
with entwined charge, spin, and superconducting orders have been detected 
experimentally. The 2D HKL model has recently been applied to model
behavior, and pairing symmetry, in heavy fermion 
superconductors.\cite{Sato2001,Xavier2008,Liu2014} 
This new 1D$_c$$\times$1D$_s$ model of coupled 1D charge ($c$) 
wires and 1D spin ($s$) chains comprises a type of system with frustrating interactions.
1D conductors have thoroughly studied  
charge- and spin-density wave instabilities at $2k_F$ and $4k_F$ respectively. Are
these instabilities overcome by the Kondo coupling, which results in coupled 1D
conductors and hence a 2D system?
The AFM ordering tendencies of the Heisenberg chain, which never actually attains long
range order, is correlated with neighboring chains by the Kondo coupling.
A plausible mean field picture might be that Heisenberg spin chains coupled, either
in phase or out of phase, with neighboring spins chains by the Kondo coupling through
the conducting wires. The corresponding spin-exchange and magnetic ordering could
introduce superstructure into the fermionic susceptibility at commensurate wavevectors,
possibly influencing the diverging susceptibility at 
2$k_F$ on the conducting systems, through the induced alternating ``magnetic field" 
transferred through the Kondo coupling.
This system may lead to a new form of magnetic polarons, accounting for the bad conductor
behavior that is so prominent in low carrier density transition metal oxides.  
Superconducting pairing tendencies become an attractive possibility, due to the 
analogies with cuprates; note that 
Nb$_{12}$O$_{29}$ is a magnetic 2D transition metal oxide with $\sim$0.1
carrier per metal ion.  We expect that low temperature and high field studies of
this system could reveal unusual transport, thermodynamic, and spectroscopic behavior.

\section{Summary}
In this electronic structure study of $o$-Nb$_{12}$O$_{29}$, two remarkable
features have been uncovered. First, regardless of whether strong on-site interaction
effects are included in the calculation, a remarkably flat Nb $4d$ band is found to
exist within a bands of several itinerant (dispersive) Nb $4d$ bands. Secondly,
when correlation effects are included, the majority spin component of this flat
band state is fully occupied, but the orbital is shared by two Nb sites, the Nb5 sites
interior to the $3\times 4$ columnar array of NbO$_6$ octahedra. The itinerant,
conducting states are found to reside along the edges of the columns, forming 1D
wires. The magnetic spins are (i) coupled to neighboring spins along a single
direction, and (ii) coupled to the itinerant carriers by
Kondo coupling. This combination appears to be quite delicate: while the orthorhombic
polytype that we have studied does not order down to 2K, the monoclinic polytype
with many structural similarities ``orders" in the sense of a 1D Heisenberg
chain, with the susceptibility following the Bonner-Fisher form.

The picture of Nb$_{12}$O$_{29}$ that arises makes more explicit 
the earlier picture of ``one magnetic Nb ion and one conducting Nb ion.'' A local
moment does emerge as required by the susceptibility data, but it is a highly
unusual {\it dimer moment} rather than a single ion moment.  Similarly, the conducting
state arises not from a single Nb ion providing an itinerant electron; instead
the conducting electron is shared between two conducting ``wires'' on either edge
of the $3\times 4$ columns of NbO$_6$ octahedra. 
The novel state that results is an illustration of how
structural complexity can give rise to new emergent phases of matter.  
This new understanding of this Nb suboxide should serve to guide study of the
more complex Nb suboxides mentioned in the Introduction.

\section{Acknowledgments}
W.E.P. acknowledges many productive interactions with K. Koepernik, M. Richter,
U. Nitzsche, and H. Eschrig during a sabbatical stay at IFW Dresden in 2006.
We also appreciate comments on the new physics of the Heisenberg-Kondo lattice model
from R. R. P. Singh, R. T. Scalettar, and W. Hu.
This research was supported by National Research Foundation of Korea
Grant No. NRF-2013R1A1A2A10008946 (K.W.L.),
by U.S. National Science Foundation Grant DMR-1207622-0 (K.W.L.)
and by U.S. Department of Energy Grant DE-FG02-04ER46111 (W.E.P.).

\end{document}